\documentclass[10pt]{amsart}

\usepackage{setspace}

\usepackage{graphicx, wrapfig, pinlabel}

\def\Z{\mathbb{Z}}

\def\R{\mathbb{R}}

\newcommand{\Cech}{\v{C}ech\,}

\theoremstyle{definition}

\begin{document}
\address{Microsoft Station Q, University of California, Santa Barbara, CA 93106}
\email{michaelf@microsoft.com}
\title{Kernel$(J)$ warns of false vacua}
\author{Michael H. Freedman}
\maketitle
\begin{abstract}
J.H.C. Whitehead defined a map $J_r:\pi_r(SO)\rightarrow \pi_r^s$ from the homotopy of the special orthogonal group to the stable homotopy of spheres. Within a toy model we show how the known computation for kernel$(J)$ leads to nonlinear $\sigma$-models with spherical source (space) and spherical target which admit false vacua separated from the true vacuum by an energy barrier. In this construction, the dimension of space must be at least $8$ and the dimension of the $\sigma$-model target at least $5$.
\end{abstract}
\onehalfspacing
\section{Introduction}
Homotopy theory and, in particular, $K$-theory have recently played a prominent role in both condensed matter (CM) and high energy physics \cite{Kane, Kitaev, RSFL, Horava, MM}. In CM applications, the discussion can often be reduced to the study of nonlinear $\sigma$-models with symmetric space targets \cite{ludwig}. In physical applications \cite{Fr}, there may be a potential function $V$ on the target symmetric space which can lead to distinct phases separated by an energy barrier. The purpose of this paper is to raise the possibility that even when there is {\it no} potential on the target symmetric space $X$, and the physical space $B$ is as simple as possible --- say a sphere or Euclidean space --- the energy landscape on (smooth) maps $\mathcal{M}(B,X)$, maps $f$ from the physical space $B$ to $X$, may have local minima, effectively creating sectors of false vacua. Similarly, the space-time action of any field $F:B\times [0,1]\rightarrow X$ interpolating between distinct vacua may be large compared to fields $F'$ remaining within a single vacuum sector. Such an $F$ is a novel kind of instanton protected by the intrinsic inefficiency of {\it any} homotopy between distinct vacua. This may be contrasted with conventional instantons, for example in Yang-Mills theory, where the instanton results from a {\it willfully} inefficient null homotopy: $$F:S^3\times[0,1]\rightarrow S^4\cong HP^1\subset HP^\infty,$$ where $F(S^3\times 0)=F(S^3\times 1)=*$, is {\it chosen} to have nonzero degree.

I will explain why no such minima are expected to arise when dim$\,B$, dim$\,X\leq 2$. After this, as the dimension increases, little is known until we meet the example, $f_0$, defined after equation (3). The focus of this paper is the surprisingly complex geometry necessarily associated to {\it any} null homotopy of: $f_0:B\rightarrow X$, where $B=S^8$ and $X=S^5$, where $f_0$ is a specific field on an $8$-dimensional spherical space taking values in the $5$-dimensional sphere. Of course, there is the question of what energy functional to use for a general $f:S^8\rightarrow S^5$. Something like $$E(f)=m^{(n-9)}\hbar\int_{S^3}d^8x\sqrt{g}\,|\nabla f|^n,\;\;8\leq n<\infty,$$ where we have chosen units with $c=1$ and $m$ a constant with units of length, is a candidate (note that for $n<8$, rescaling $f$ toward a point $x\in S^8$ by precomposing with a conformal transformation with $x$ a repelling fixed point makes $E$ approach zero). Another possibility related to the choice $n=\infty$ is to let $E$ be the Lipschitz constant of $f$ (times a unit of energy). These choices entail serious analytical difficulties which are circumvented here by choosing a surrogate energy \begin{equation}E_{\rm top}=\max_{x\in S^5}\sum_{i=0}^7 b_i\big{(}f^{-1}(x)\big{)},\end{equation} $f$ is presumed to be a smooth ($C^\infty$) function and $b_i$ the $i^{\rm th}$ Betti number, $b_i={\rm rank}\; \check{H}^*(X;R)$ (I have now dropped coupling constants and reference to units). We use the \Cech cohomology $\check{H}^i$ because it is appropriate for the nonregular point inverse images, which may be arbitrary closed subsets of $B$.

Similarly, define the action $S_{\rm top}$ of a field $F$ on space-time $F:=f_t:S^8 \times [0,1]\rightarrow S^5$ by \begin{equation} S_{\rm top}(F)=\max_{x\in S^5}\sum_{i=0}^7 b_i\big{(}F^{-1}(x)\big{)}.\end{equation}
Both $E_{\rm top}$ and $S_{\rm top}$ are nonlocal --- that is not integrated up from a locally defined quantity --- but nevertheless seem to capture an essential feature of more physical Hamiltonians and Lagrangians, respectively: maps of high energy (action) are generally those with complicated point preimages (see {\small \sc Figure 1}). With this definition, $f_0$ has $E_{\rm top}(f_0)=2$ and can be deformed to the consant map $f_1$, that is the true vacuum, with $E_{\rm top}(f_1)=1$, but during the deformation we find that $E_{\rm top}(f_{t_0})>2$ for some $t_0\in (0,1)$. Thus $E_{\rm top}(f_{t_0})$ is an energy barrier between a false vacuum $f_0$ and the true vacuum. Similarly, the action must be large for any $F$ interpolating between $f_0$ and $f_1$: $S_{\rm top}(F)\geq 23$ whereas for paths staying within a vacua, for example $S_{\rm top}\bigg{(}(F(y,t)):=f_0(y)\bigg{)}=2$ and $S_{\rm top}\bigg{(}(F(y,t)):=f_1(y)\bigg{)}=1$, the action can be much smaller. I should emphasize that although topology is used to define $E_{\rm top}$ and $S_{\rm top}$, these quantities are {\it not} homotopy invariants: as $f_0$ is deformed, there will be times when $E_{\rm top}(f_t)$ jumps; similarly for $S_{\rm top}(F)$ if $F$ were also deformed. What $E_{\rm top}$ and $S_{\rm top}$ {\it do} depend on is the topological complexity of the inverse images of points $x\in S^5$, and these can jump whenever $x$ is (transiently) a singular value.

So one may understand ({\small\sc Figure} 1) how preimage complexity roughly encodes energy. The surprise is that it may be necessary to increase the energy en route to decreasing it. It would be as if in passing from {\small\sc Figure} 1(a) to {\small\sc Figure} 1(c) we had to go through {\small\sc Figure} 1(b).

\begin{figure}[htpb]
\labellist \normalsize\hair 2pt

   \pinlabel $(a)$ at 10 -10
   \pinlabel $(b)$ at 222 -10
   \pinlabel $(c)$ at 434 -10
   \pinlabel $\text{high energy}$ at 95 -12
   \pinlabel $\text{low energy}$ at 520 -12
   \pinlabel $\text{very high energy}$ at 310 -12
   \pinlabel $S_{\rm top}=13$ at 90 119
   \pinlabel $S_{\rm top}=3$ at 520 119

\endlabellist
\centering
\includegraphics[scale=0.6]{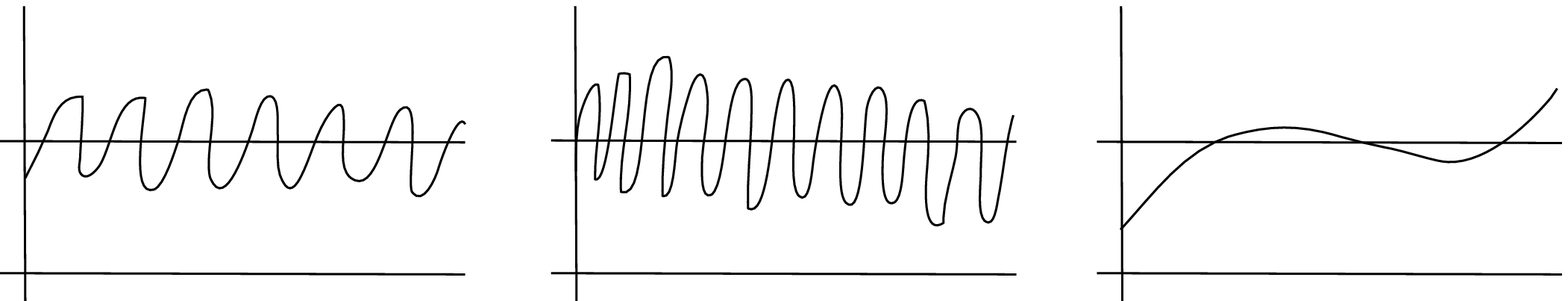}
\caption{} \label{fig_states}
\end{figure}

If we distance ourselves from energy functionals, the basic idea of needing to ``pass over a saddle'' on the way to lowering complexity can easily be illustrated using immersions. The path components of immersions $\mathcal{I}(S^1,\R^2)$ are known to be indexed by the winding number in $\mathbb{Z}$. It is easy ({\small\sc Figure} 2) to find two immersions $\alpha_0$ and $\alpha_1$ with equal winding where the ``complexity'' as measured by the number of multiple points, must increase during any regular homotopy $\alpha_t$, $t\in [0,1]$, beyond the initial or final values.

\begin{figure}[htpb]
\labellist \normalsize\hair 2pt

   \pinlabel $\alpha_0$ at 38
 -10
  \pinlabel $\alpha_{1/4}$ at 139 -10
   \pinlabel $\alpha_{1/2}$ at 245 -10
   \pinlabel $\alpha_{3/4}$ at 360 -10
\pinlabel $\alpha_{1}$ at 457 -10

\endlabellist
\centering
\includegraphics[scale=0.8]{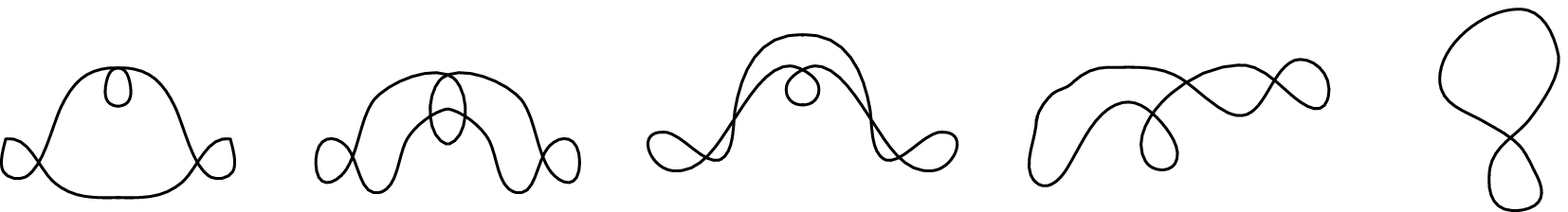}
\caption{} \label{fig_states}
\end{figure}

Unlike $E_{\rm top}$, the complexity in this example seems, in spirit, opposite to conventional energy functionals: for example, a constant map, though not an immersion, would have infinite multiplicity and hence high, not low, energy.

It is hoped that some insight can be gained even from the ``toy functionals'' $E_{\rm top}$ and $S_{\rm top}$. The key to the high dimensional example is the noninjectivity of Whitehead's $J$-homomorphism from homotopy theory.

In summary, this paper is a message from topology to physics that fields {\it might} become trapped in local minima for the energy of a nonlinear $\sigma$-model, even when the space-time and the target space are as homogeneous as possible: both spheres. In our example, the dimensions are a bit high (except for string theorists) and the action is a toy. The challenge for the reader is to determine if this same phenomenon can occur, or contrarywise is excluded, in more conventional, and perhaps lower dimensional, physical situations.

\section{The Example}
In homotopy theory, the $J$-homomorphism is the map $J_r:\pi_r(SO)\rightarrow\pi_r^s$ from the homotopy of the special orthogonal group to the $r^{\rm th}$ stable stem $\pi_r^s(:=\pi_{r+d}(S^d)$ for $d>r+1$). We can interpret $\alpha\in\pi_r(SO)$ as a (stable) normal framing of the $r$-sphere $S^r$ and $\beta\in\pi_r^s$ can, by the Pontrjagin-Thom construction, be encoded as a normally framed $r$-manifold $M^r$. The Pontrjagin-Thom construction associates to $\beta:S^{r+d}\rightarrow S^d$, the preimage of a regular value $*\in S^d$, $M^r:=\beta^{-1}(*)$. The normal bundle of $M^r$ in $S^{r+d}$ is framed by pulling back the normal framing of $*$ in $S^d$. For $d>r+1$, the stable normal cobordism class of $M^r$ is equivalent to the homotopy class of $\beta$. In these terms, the $J$-homomorphism simply includes spheres (with framed normal bundle) into manifolds (with framed normal bundle). It is known that $J_r$ is injective unless $r=4n-1$, $n\geq 1$, and is the epimorphism $J_r:\Z\rightarrow \Z_{4n/B_{2n}}$, where $B_{2n}$ is the $2n^{th}$ Bernoulli denominator. The first relevant images of $J$ are Im$J_3=\Z_{24}$, Im$J_7=\Z_{240}$, and Im$J_9=\Z_{504}$. The discussion requires only that $J_r$ have a kernel, so there are examples of exotic local minima for the obvious dimensional extension\footnote{Extend the definition of $E_{\rm top}$ and $S_{\rm top}$ given in equations (1) and (2) by extending the range of the summations to $({\rm spatial \;dimension})-1$.} of $E_{\rm top}$ (and $S_{\rm top}$ after crossing the source with $[0,1]$) to maps $S^{2(4n-1)+2+i}\rightarrow S^{(4n-1)+2+i}$ for all $n>0$ and $i\geq 0$. Increasing $n$ means going to a higher stable stem, and increasing $i$ merely means suspending maps within that stable stem. I analyze the case $n=1$ and $i=0$. Nothing changes as $i$ increases, however with increasing $n$ the characteristic class computations are slightly different and more significantly a refinement coming from Seiberg-Witten theory is not present. In all cases, however, there is a similar energy barrier.

The homotopy group $\pi_3(SO)\cong \Z$ is generated by the composition \begin{equation}g:S^3\cong SU(2)\xrightarrow[]{\rm double\; cover}SO(3)\hookrightarrow SO.\end{equation} $24g$ is obtained by precomposing $g$ with $(\phi,\theta_1;\theta_2)\mapsto (24\phi,\theta_1;\theta_2)$ in spherical coordinates. The map we study $f_0:S^8\rightarrow S^5$ may be defined by writing $S^8=S^3\wedge S^4$, where the join sumbol $\wedge$ indicates the space of line segments ``joining'' the two factors: $P\wedge Q:=P\times Q\times [0,1]/(p,q,0)=(p,q',0)$ and $(p,q,1)\equiv (p',q,1)$, for all $p,p'\in P$ and all $q,q'\in Q$. In these coordinates, $f_0(S^3)=+$ and $f_0(S^4)=-$, $+(-)$ the north (south) pole of $S^5$. The normal direction to $S^3$ in the join consists of a unit $5$-disk $D_s^5$ at every point $s\in S^3$ and $f_0$ wraps this $D_s^5$ degree = one over $S^5$ by exponentiation. The only ambiguity is that we have not yet said how $D^5_s$ is identified with the unit disk in the target space $T_+(S^5)$ as a function of $s$. When $s=*\in S^3$, the base point, say the identity of $SU(2)$, the identification is an arbitrary isometry. For general $s$, precompose this arbitrary identification with $24g$ as defined above, where $SO$ acts on $\R^5$ through the intermediate $SO(5)$, $SO(3)\subset SO(5)\subset SO$.

By inspecton, every point preimage of $f_0$ is a $3$-sphere except the preimage of the south pole, which is a $4$-sphere. Since $${\rm rank}\;\check{H}^i(S^n;R) = \begin{cases}

  1, & \text{for $i=0$ or $n$,} \\

  0, & \text{otherwise},
\end{cases}$$ we see that $S_{\rm top}(f_0)=2$.

Since it is known that $J(24g)\simeq f_0$ is homotopically trivial, $f_0$ is homotopic to $f_1$, the constant map $S^8\rightarrow S^5$, taking each point of $S^8$ to the south pole $-$ of $S^5$. The only nonempty preimage of $f_1$ is $f_1^{-1}(-)=S^8$, so $E_{\rm top}(f_1)=\sum_{i=0}^7 b_i=1$. We now state: \\

\noindent\textbf{Theorem 1}. {\it If $f_t$ is a smooth family, $F:=f_t:S^8\times[0,1]\rightarrow S^5$ with $f_1$ constant, then $S_{\rm top}(F)\geq 23$.}
\begin{proof}
By Sard's theorem, the regular values on $S^5$ are an open dense set of full Lebesgue measure; let $*\in S^5\setminus (-)$ be one. By the Inverse Function Theorem, $F^{-1}(x)$ is a smooth, normally framed $4$-dimensional manifold $M\subset S^8\times[0,1]$, with $\partial M=S^3\subset S^8\times 0$. Cap $M$ off with a $4$-disk to obtain a smooth, closed $4$-maifold $\hat{M}$. Since $\hat{M}$ has a normal framing away from a point, the normal characteristic classes $w_1(\nu_{\hat{M}})=w_2(\nu_{\hat{M}})=0$. By the Whitney sum formula those tangential classes also vanish $w_1(\tau_{\hat{M}})=w_2(\tau_{\hat{M}})=0$, so $\hat{M}$ is spin. The first obstruction to trivializing the normal frame bundle over $\hat{M}$ is $24\gamma\in H^4(\hat{M};\pi_3(SO))\cong \Z$, where $\gamma$ is a generator, $\gamma=\pm 1\in\Z$. The first Pontrjagin class $p_1(\tau_{\hat{M}})$ is twice this obstruction, $p_1(\tau_{\hat{M}})=\pm 48\gamma$. On the other hand, in dimension $4$ the Hirzebruch $L$-genus reduces to: \begin{equation} \text{signature of cup-product on }H_2(\hat{M};\Z)/{\rm torsion}=:\sigma(\hat{M})=\frac{p_1(\hat{M})}{3},\end{equation} so $\sigma(\hat{M})=\pm 16$. The |signature| is a lower bound on the second Betti number $b_2(\hat{M})$. But, by excision, $b_2(M)=b_2(\hat{M})$. Putting this together we find \begin{equation} b_2(M)\geq 16\end{equation} since $b_0(M)\geq 1$. We may conclude that \begin{equation}\sum_{i=0}^7 b_i(M)\geq 17.\end{equation} Since $S_{\rm top}$ is defined as the maximal such sum over point preimages, $S_{\rm top}(F)\geq 17$.

Because $\hat{M}$ is smooth, spin, and $4$-dimensional (since we are considering kernel $J_3$), there is more refined information on the lower bound to $b_2$ available from the Seiberg-Witten equations. It is known that $\hat{M}$ must have $b_2$ as large as that of the $K3$ surface, $b_2(\hat{M})\geq 22$. This follows from Furuta's ``$10/8$-theorem'' \cite{Fu}, and also from previous unpublished work of Peter Kronheimer. The estimate $b_2(\hat{M})\geq 22$ is Donaldson's ``Theorem C'' in the special case $\pi_1(\hat{M})=0$. This estimate implies $S_{\rm top}(F)\geq 23$.
\end{proof}

This is the Lagrangian result. The corresponding energy barrier is identified by the next theorem.\\

\noindent\textbf{Theorem 2}. {\it If $f_t$ is a smooth family $F:=f_t:S^8\times[0,1]\rightarrow S^5$ with $f_0$ as defined and $f_1$ constant, then there exists a $t\in(0,1)$ where $E_{\rm top}(f_t)\geq 3$. Recall $E_{\rm top}(f_0)=2$ and $E_{\rm top}(f_1)=1$.}
\begin{proof}
Begin with the same $(M,\partial M)$, $\partial M=S^3$, as in the proof of Theorem 1, except now restrict $M$ to be only the connected component containing $\partial M$. Let $h:M\rightarrow [0,1]$ be the inclusion $M\subset S^8\times [0,1]$ followed by projection to the second factor. Intuitively, we may think of $h$ as a Morse function, but this cannot in fact be assumed. All we really know is $h^{-1}(0)=\partial M\cong S^3$, $h^{-1}(1)=\O$, and $h$ is smooth. A lower bound to $$b:=\max_{t\in[0,1]}\sum_{i=1}^7 b_i(h^{-1}(t))$$ is the lower bound to the barrier $\max_{t\in[0,1]}E_{\rm top}(f_t)$ obtained by restricting attention to preimages of $*\in S^5$. We show $b\geq 3$ by assuming for a contradiction that $b=2$. Let $x_0\in [0,1]$ be the maximum value for which $h^{-1}(x)\neq\O$. Since we assumed $M$ connected, for all $x\in [0,x_0]$ we have $h^{-1}(x)\neq \O$. Let $x_1\in (0,x_0)$ be any regular value for $h$. $h^{-1}(x_1)$ is a smooth, closed $3$-manifold so by the assumption $b=2$, $h^{-1}(x_1)$ must be a real (equivalent rational) homology $3$-sphere (otherwise its first and second Betti numbers would contribute to the sum).

A key property of \Cech cohomology is that it commutes with inverse limits. This will now be used twice.\\

\noindent\textbf{Lemma 3}. {\it For all $x_1,x_2$ $h$-regular values in $[0,x_0]$, $h^{-1}[x_1,x_2]$ must have the real \Cech cohomology of $S^3$.}
\begin{proof}
Suppose this fails for some $[x_1,x_2]$ pick an $h$-regular $x_3\in (x_1,x_2)$ near $(x_1+x_2)/2$ and write $$h^{-1}[x_1,x_2]=h^{-1}[x_1,x_3]\bigcup_{h^{-1}(x_3)} h^{-1}[x_3,x_2].$$ Since $h^{-1}(x_3)$ is normally collared ($x_3$ is an $h$-regular value) we may use the usual Mayer-Vietoris sequence to conclude that either $h^{-1}[x_1,x_3]$ or $h^{-1}[x_3,x_2]$ fails to have the real \Cech cohomology of $S^3$ and the additional homology $\check{H}^i(h^{-1}[x_1,x_3];\R)$, $i=1,2$ restricts from $\check{H}^i(h^{-1}[x_1,x_2];\R)$ (for notational simplicity, write $h^{-1}[x_1,x_3]$ where $h^{-1}[x_3,x_2]$ could instead occur).

Picking a regular $x_4\in (x_1,x_3)$ and near $(x_1+x_3)/2$, we may conclude similarly that one of the pieces, say $h^{-1}[x_1,x_4]$, is not a real \Cech cohomology $S^3$. Proceeding in this way, find a sequence of nested intervals $I_j$ each no more than, say, $60\%$ the length of the previous with the cohomology of $h^{-1}(I_j)$ containing a nontrivial subspace which restricts from $\check{H}^i(h^{-1}[x_1,x_2];\R)$, $i=1,2$. Taking inverse limits by letting $z=\cap_{j=1}^\infty I_j$, we find $\check{H}^i(h^{-1}(z);\R)$, $i=1,2$, also contains a nontrivial subspace restricting from $\check{H}^i(h^{-1}[x_1,x_2])$, contradicting $b=2$.
\end{proof}
From the lemma, and once again applying the Mayer-Vietoris sequence to decompositions inverse to $h$-regular values, we conclude:\\

\noindent\textbf{Lemma 4}. {\it For every regular value $x_1\in [0,x_0]$, the natural map $$\check{H}^*(M;\R)\rightarrow\check{H}^*(h^{-1}[x_1,x_0];\R)$$ is an isomorphism.}
\begin{proof}
Write $$M=h^{-1}[x_1,x_0]\bigcup_{h^{-1}(x_1)} h^{-1}[0,x_1]$$ and use Lemma 3 to recognize the second piece as a cohomology collar. Apply Mayer-Vietoris along $h^{-1}(x_1)$.
\end{proof}
The proof of Theorem 2 is now completed by a second passage to inverse limits. Write $$h^{-1}(x_0)=\bigcap_{h-{\rm regular}\;x_{1_i}}h^{-1}[x_{1_i},x_0].$$ By Lemma 4, the cohomology of each $h^{-1}[x_{1_i},x_0]$ is given isomorphically by restriction from $\check{H}^*(M;\R)$, so passing to limits, $\check{H}^*(h^{-1}(x_0);\R)\cong \check{H}^*(M;\R)$ via inc$^*$. But since $b_2(M)\geq 22$, this implies $b_2(h^{-1}(x_0))\geq 16$. Thus $2=b\geq b_2(h^{-1}(x_0))+b_0(h^{-1}(x_0))\geq 22+1=23$, a contradiction.
\end{proof}
\section{Low Dimensions}
The spaces of maps, $\mathcal{M}(S^1,S^1)$, $\mathcal{M}(S^2,S^1)$, and $\mathcal{M}(S^2,S^2)$ probably do not hold any surprises through in the last case there are some subtleties.

When the target is $S^1$, ``circular'' Morse theory can be used to steadily simplify maps until they are constant or in the $S^1\rightarrow S^1$ case, possibly a covering map. If we try to proceed to the case of maps $S^3\rightarrow S^1$, $\mathcal{M}(S^3,S^1)$, unknown issues are encountered. We may lift each $f:S^3\rightarrow S^1$ to $\tilde{f}:S^3\rightarrow\R$. If $\tilde{f}$ happens to be a ``self-indexing'' (critical values of higher index are larger) a theorem of Waldhausen \cite{W} shows that $\tilde{f}$ can be monotonely simplified to the standard height function. But the existence of such monotone simplifications appears to be open if $\tilde{f}$ is not self-indexing. If no simplifying path from $\tilde{f}$ is monotone, then the passage between a local minimum $\tilde{f}$ and global minimum would be analogous to the instanton we produced between the vacua on $S^8$.

$\mathcal{M}(S^2,S^2)$ is quite interesting. I would conjecture that there are no energy barriers: that for reasonable choices of energy topological or otherwise, energy may be monotonically reduced until a cyclic branch cover is reached. This appears to be true for $E_{\rm top}$, unsigned area, and harmonic map energy, $$\int_{S^2}\;d^2x\,|\nabla f|^2.$$ In fact, work of Topping \cite{T} on harmonic map flow might be used in verifying such conjectures. The essential point is that because the target $S^2$ is {\it positively curved}, the local energy density $|du|^2$ is not bounded by the usual Eells-Sampson maximal principle but instead may exhibit a specific singularity called ``bubbling'' under harmonic map flow. Intuitively, according to Topping, one should be able to run harmonic map flow until just before a singularity and then pause to do a $\pi$-twist to prevent the flow from being caught on the singularity and then resume the flow. The phenomenon of bubbling and the flow catching on a singularity are certainly intrinsic since the Hopf map $S^3\rightarrow S^2$ can be regarded as an ``essential loop of degree zero maps $S^2\rightarrow S^2$''; only the flow catching the singularity prevents the loop from being contracted.

Proceeding by dimension, the techniques of \cite{BF} and \cite{DGS} might be useful in bounding the height of energy barriers within $\mathcal{M}(S^3,S^2)$ if they in fact exist, but so far nothing is known.

For $\mathcal{M}(S^1,S^3)$ (and, more generally, $\mathcal{M}(S^m,S^{m+2})$ where $m\geq 1$), the phenomenon of knotting implies the existence of $E_{\rm top}$ energy barriers. Furthermore, there is a physically interesting conformally invariant energy associated to $\mathcal{M}(S^1,S^3)$, \cite{FHW}, but it is unknown if it has local (but not global) minima within a given knot type. The gradient flow of this functional has been studied as a potential ``unknotting'' algorithm.

\section{Energy Barriers from Mathematical Logic}
One surprising geometric output of mathematical logic is the existence of enormous energy barriers related to recognition problems. It was shown in the 1950's by Boone and Novikov that the triviality problem for finitely presented groups is undecideable. However the Tietze theorem tells us that any two finite group presentations of the same group are joined by a finite string of four simple moves on presentations. The only way these two statements can be compatible is:\\

\noindent\textbf{Fact 5}. {\it The minimal number of Tietze moves required to reduce a presentation of the trivial group of total size $n$ to the empty presentation must be a function $f(n)$ which grows more rapidly than any recursive function.}\\

This fact finds an even more geometric echo in work of Nabutosky and Weinberger \cite{NW}. An example is:\\

\noindent\textbf{Fact 6}. {\it If one considers flat piecewise linear imbedding $e:S^5\hookrightarrow \R^6$ (or one may take $S^d\subset \R^{d+1}$, for and $d\geq 5$) consisting of $n$ top dimensional simplices, then the degree of refinement $f(n)$, which might be required before there is a piecewise linear isotopy to a {\it convex} imbedding $e_{\rm convex}:S^5\hookrightarrow \R^6$, also grows faster than any recursive function.}\\

The point is that any recursive growth rate would lead to an algorithm for recognizing $5$-spheres, which is impossible. In attempting to recognize $S^5$, the chief difficulty is determining the triviality of the fundamental group.

\section{Summary}
We have shown in a toy model that even for nonlinear $\sigma$-models whose range and domains are round spheres, false vacua and energy barriers can arise in the space of maps. The tool is J.H.C. Whitehead's $J$-homomorphism, $J_r:\pi_r(SO)\rightarrow \pi_r^s$, from the homotopy of the special orthogonal group to the stable homotopy of spheres. Energy barriers and the instantons which cross them have diverse mathematical origins. Here the kernel of $J$ has been used to describe a new class of barrier and instanton.

{\large November 8, 2011}
\end{document}